\documentclass[pra,aps,twocolumn,showpacs,superscriptaddress]{revtex4-1}
\usepackage{color,colordvi,graphicx,amssymb,epsfig,amsmath}

\usepackage[utf8]{inputenc}
\RequirePackage{amsfonts}
\usepackage{epstopdf}

\newcommand{\be}{\begin{equation}}
\newcommand{\ee}{\end{equation}}
\newcommand{\bea}{\begin{eqnarray}}
\newcommand{\eea}{\end{eqnarray}}

\begin{document}
\title{Phases, many-body entropy measures and coherence of interacting bosons in optical lattices}

\author{R. Roy}
\affiliation{Department of Physics, Presidency University, 86/1 College Street, Kolkata 700 073, India.}
\author{A. Gammal}
\affiliation{Instituto de Fisica, Universidade de S\~{a}o Paulo, CEP 05580-090, S\~{a}o Paulo, Brazil }
\author{M. C. Tsatsos}
\affiliation{Instituto de F\'sica de S\~ao Carlos, Universidade de S\~ao Paulo,
CP 369,13560-970, S\~ao Carlos, SP, Brasil}
\author{B. Chatterjee}
\affiliation{Department of Physics, Indian Institute of Technology Kanpur, India}
\author{B. Chakrabarti}
\affiliation{Instituto de Fisica, Universidade de S\~{a}o Paulo, CEP 05580-090, S\~{a}o Paulo, Brazil }
\affiliation{Department of Physics, Presidency University, 86/1 College Street, Kolkata 700 073, India}
\author{A. U. J. Lode}
\affiliation{Wolfgang Pauli Institute c/o Faculty of Mathematics,
University of Vienna, Oskar-Morgenstern Platz 1, 1090 Vienna, Austria}
\affiliation{Vienna Center for Quantum Science and Technology,
Atominstitut, TU Wien, Stadionallee 2, 1020 Vienna, Austria}
\affiliation{Department of Physics, University of Basel, 4056 Basel, Switzerland}

\date{\today}
 
\begin{abstract}
Already a few bosons with contact interparticle interactions in small optical lattices feature a variety of quantum phases: superfluid, Mott-insulator and fermionized Tonks gases can be probed in such systems. 
To detect these phases -- pivotal for both experiment and theory -- as well as their many-body properties we analyze several distinct measures for the one-body and many-body Shannon information entropies. We exemplify the connection of these entropies with spatial correlations in the many-body state by contrasting them to the Glauber normalized correlation functions. To obtain the ground-state for lattices with commensurate filling (i.e. an integer number of particles per site) for the full range of repulsive interparticle interactions we utilize the multiconfigurational time-dependent Hartree method for bosons (MCTDHB) in order to solve the many-boson Schr\"odinger equation. We demonstrate that all emergent phases -- the superfluid, the Mott insulator, and the fermionized gas can be characterized equivalently by our many-body entropy measures and by Glauber's normalized correlation functions. In contrast to our many-body entropy measures, single-particle entropy cannot capture these transitions. 
\end{abstract}

\pacs{}
\keywords{ultracold atoms, many-body quantum mechanics}
\maketitle

\section{Introduction}
Ultracold atoms provide a testing ground for many-body physics~\cite{Bloch}. Utilizing Feshbach resonance management, the interatomic interaction can be tuned to any desired value~\cite{interactions}. This tunability allows the production of weakly or strongly correlated Bose Einstein condensates (BECs). A variety of different geometries and topologies is realizable by suitably manipulating the magneto-optical trapping potentials~\cite{arbpot}. For instance, interacting bosons in optical lattices, that is, a spatially periodic potential, have been shown to exhibit a quantum phase transition from a superfluid state (SF) to a Mott insulator (MI)~\cite{Jaksch,Greiner,Haller}. 

One-dimensional optical lattices loaded with strongly interacting bosons have been shown to be experimentally more challenging than their three-dimensional counterparts: quantum fluctuations are not negligible~\cite{haldane}, trigger correlations \cite{haller1,kinoshita} and are the focus of the present paper. 

The zero-temperature SF to MI transition is commonly described by the Bose-Hubbard model~\cite{Fisher}. For weak interparticle interactions, a Bose gas in an optical lattice of moderate depth is in the superfluid phase: the many-body state features long-range coherence. This SF phase persists, as long as the interatomic interaction is small compared to the tunneling coupling -- i.e. the parameter of the Bose-Hubbard Hamiltonian that determines the lattice depth and hence tunneling between neighboring sites. The SF can be characterized as a state in which all particles are delocalized between all sites of the lattice. When the repulsive interaction is large compared to the tunneling coupling, each site is filled up with an identical number of bosons and the coherence between different sites is completely lost: the many-body state enters the Mott-insulating phase. 

Even though the Bose-Hubbard model can aptly describe the above SF to MI transition in lattices, its regime of validity is restricted to the case where site-localized Wannier states are an appropriate basis set~\cite{Grigory}. It has been demonstrated that a general quantum many-body description, valid at all interaction strengths, is necessary for the regime beyond the Bose-Hubbard model~\cite{Sakmann2009}. One such theory is the  the multiconfigurational time-dependent Hartree (MCTDH) method \cite{mctdh}. In Ref.~\cite{Brouzos} MCTDH was applied to small lattices and it was found that correlations depend both on the commensurability and the strength of the interparticle interactions. In the present paper, we solve the full many-body Schr\"odinger equation at a high level of accuracy by using the multiconfigurational time-dependent Hartree method for bosons (MCTDHB)~\cite{Ofir,OfirJCP} implemented in the MCTDH-X software~\cite{axel1,axel2,ultracold}. 
The motivation for our present work is to investigate the pathway from superfluid through the Mott insulating to the fermionized phase and explain it in terms of production of many-body information entropy.

Entropies of quantum systems were long ago introduced and used as a measure of the degree of order of a given quantum state~\cite{Massen1,Massen2,Massen3}. Entropy measures for many-body states were shown to saturate in time to the values given by estimates of Gaussian orthogonal ensembles (GOE) of random matrices for the time evolution triggered by an interaction quench~\cite{SIE_1}. This situation, when different entropy measures take on the GOE values, is referred to as \emph{statistical relaxation} and was shown to also reflect in the correlation functions of the many-body state~\cite{SIE_1}. 

Below, we analyze this connection of many-body entropies and correlation functions for the quantum phases of many-body states in an optical lattice, commensurately filled with bosons that interact with a repulsive contact interaction. We characterize the many-body state by calculating the produced entropy and the spatial correlations as a function of the interparticle interaction strength. We demonstrate that the production of different types of entropies -- defined below -- and the correlation functions can be used to identify the phase of the system.

The \emph{Shannon information entropy (SIE)} can be computed from the one-body density, i.e., the diagonal of the first-order reduced density matrix, in coordinate or momentum space~\cite{Guevara1,Guevara2,Sagar,Sen}. In this case, SIE is a measure of the spatial delocalization of the one-body probability distribution of the system. Since SIE of the one-body density is computed from a \emph{single-particle quantity}, it is insensitive to the many-body physics of the considered system and its correlations.

In this work, in order to remedy this shortcoming of the SIE of one-body quantities, we generalize the concept and introduce SIEs on the many-body level. In particular, we define and investigate: $a)$ the \emph{two-body Shannon information entropy} calculated from the second-order reduced density matrix, $b)$ the \emph{occupation SIE (O-SIE)}, i.e. the entropy of the eigenvalues of the reduced first-order density matrix~\cite{Brezinova2012,Tsatsos2015,SIE_1} and $c)$ the \emph{Shannon information entropy of the coefficients (C-SIE)} of the expansion of the state on some many-body basis (see Sec.\ref{Hamiltonian})~\cite{SIE_1}. These entropies -- as opposed to the single-particle SIE -- directly relate to the many-body physics of the system. The O-SIE and C-SIE are many-body measures that allow us to get a quantitative criterion for the applicability of theories like, for instance, the Gross-Pitaevskii (GP) mean-field theory; the C-SIE and O-SIE take on extremal values for mean-field GP states while any other value heralds beyond-mean-field physics.
In the following, we study the many-body entropy production of interacting ultracold atoms in a one-dimensional optical lattice as a function of the strength of the interparticle interactions and the number of bosons per site. We establish a relation of the phase diagram to the C-SIE and O-SIE entropy measures. We complement the emergent picture with an analysis of Glauber's normalized first- and second-order correlation functions~\cite{RJG}. Refs.~\cite{SIE_1,Brezinova2012} have already demonstrated that a fundamental connection between the correlation functions and many-body entropy measures exists, at least for the temporal evolution following a quench. In our present study, we demonstrate that this relation between correlations and many-body entropy also holds for the ground states of bosons in optical lattices.

For small barriers and weak interactions, the system is a superfluid, aptly described within a mean-field theory and characterized by a smooth increase (decrease) of the Shannon information entropy of the one-body density in coordinate (momentum) space~\cite{Massen1,Massen2,Massen3} for increasing interaction strength.

For larger barriers or stronger interactions and one atom per site, i.e. commensurate filling, the system becomes a Mott insulator with no phase coherence between distinct sites. For this MI phase, the reduced one-body density matrix acquires multiple eigenvalues: fragmentation emerges \cite{Oliver,spekkens,mueller,lode17,PV,PNAS}. We characterize the many-body state through the superfluid-to-Mott-insulator phase transition using our many-body entropy measures.  For sufficiently strong interparticle interactions we observe a complete saturation of entropies to values predicted by the Gaussian Orthogonal Ensemble of random matrix theory~\cite{Kota2001}.

If the commensurate filling of the lattice is larger than one atom per site, a second phase transition emerges for very strong interparticle interactions: the fragmented Mott insulator transmogrifies into a so-called fermionized state~\cite{Gir,Ofir2}. The onsite repulsion forces the density of the system to develop an intrasite structure. The C-SIE and O-SIE measures of many-body information entropies also saturate for large interactions in the case of the larger filling factor with, however, no limiting value from a Gaussian orthogonal ensemble existing for this case.

We quantify the coherence properties of all three above phases -- condensed superfluid, fragmented Mott insulator and fermionized phase -- by computing Glauber's normalized first- and second-order correlation functions. We find that the classification of phases using correlation functions agrees with the classification of phases using many-body entropy measures. The fundamental relation between many-body entropies and coherence properties in ultracold bosonic atoms, found in Refs.~\cite{SIE_1,Brezinova2012} in the time evolution following a quench, is extended to the present case of stationary states in lattices.

The paper is structured as follows. In Sec.~\ref{Hamiltonian} we expose the Hamiltonian and introduce the many-body theory and our quantities of interest: Glauber correlation functions and many-body entropy measures. Section~\ref{results} deals with the results of our analysis of states for various interaction parameters and filling factors. We conclude in Sec.~\ref{Conclusion}.

\section{Theory} \label{Hamiltonian}

\subsection{Hamiltonian}
Consider a system of $N$ bosons interacting by a contact interparticle interaction potential in one spatial dimension. In dimensionless units, such a system is governed by the following Hamiltonian:
\begin{equation}
H = \sum_{i=1}^N \left( \frac{1}{2} \frac{\partial^2}{\partial x_i^2} + V(x_i) \right) + \lambda \sum_{i<j}^N \delta (x_i - x_j). \label{eq:HAM}
\end{equation}
Here, $V(x)= V_0 \sin^2(kx)$ is the lattice potential, where $k=\frac{\pi}{d}$; $V_0$ is the depth and $d$ is the periodicity of the lattice. The strength of the two-body interactions, $\lambda$, can be experimentally tuned almost at will in quasi-one-dimensional systems by manipulating the strength of the transversal confinement~\cite{1d_scattering}. In the remainder of the paper, we set a depth of $V_0 = 12.0$, $x_{\text{min}}=-4.7124$, $x_{\text{max}}=4.7124$ and $d=3$ wells are considered. The depth $V_0$ is chosen such that it allows superfluidity, for appropriately chosen boson number $N$ and interaction strength $\lambda$. We find the stationary solutions of the many-body Schr\"odinger equation with periodic boundary conditions and obtain the observables defined in Sec.~\ref{QOI} as a function of the strength of the interparticle interactions $\lambda$ and the number of atoms per lattice site.
 
\subsection{MCTDHB}
In the multiconfigurational time-dependent Hartree for bosons (MCTDHB) approach, the wave function of the interacting $N$-boson problem is expanded over a set of permanents. Permanents are symmetrized bosonic states of $N$ particles in $M$ single-particle states. Each permanent can be constructed by acting products of $N$ boson creation operators $b^\dagger_k$ ($k=1,...,M$) onto the vacuum $\vert vac \rangle$:
\begin{eqnarray}
& \vert \Psi(t) \rangle & ~=~ \sum_{\vec{n}} C_{\vec{n}}(t) \vert  \vec{n}; t \rangle; \label{cansatz} \\
& \vert \vec{n} \rangle & ~= \vert n_1,n_2, ...  n_M \rangle = \prod_{i=1}^M \left[ \frac{\left(b_i^\dagger (t)\right)^{n_{i}}}{\sqrt{n_i !}}\right] \vert vac \rangle.
\label{ansatz}
\end{eqnarray}
Here, each operator $b^\dagger_k(t)$ creates a boson occupying the time-dependent single-particle state (orbital) $\phi_{k}(x,t)$. The number of possible configurations of $N$ bosons in $M$ orbitals is equal to $\binom{N+M-1}{N}$ and defines the number of complex-valued coefficients $C_{\vec{n}}(t)$ in Eq.~\eqref{cansatz}. A formal variational treatment with the above ansatz leads to the MCTDHB equations of motion~\cite{Ofir,Streltsov2007,OfirJCP}. The solution of the latter yields the time evolution of the coefficients $C_{\vec{n}}(t)$ and orbitals $\phi_k(x,t)$ that built-up our solution: an approximation to the solution of the time-dependent Schr\"odinger equation at a desired (arbitrary) degree of accuracy. In this work, we find the eigenstates of the MCTDHB equations by propagating the MCTDHB equations in imaginary time using the MCTDH-X package~\cite{axel1,axel2,ultracold}. 
In the limit of infinite orbitals, $M\rightarrow\infty$, the set of permanents $\vert n_1,n_2, \dots  n_M; t \rangle$ in Eq.~\eqref{ansatz} spans the complete $N$-particle Fock space and MCTDHB becomes exact~\cite{exact,exact2,exact3}.

\subsection{Quantities of interest}\label{QOI}
We now introduce the quantities that we use to characterize the solutions of the MCTDHB equations.
\paragraph{\bf Entropies:}
The Shannon information entropies of the one-body density in coordinate space $\rho^{(1)}(x)=\langle \Psi \vert \hat{\Psi}(x)^\dagger \hat{\Psi}(x) \vert \Psi \rangle$ is given by
\begin{equation}
S_x(t) = - \int dx \rho^{(1)}(x,t) \ln [\rho^{(1)}(x,t)].
\label{SIE1x}
\end{equation}
Analogously, for the momentum density, $\rho^{(1)}(k)=\langle \Psi \vert \hat \Psi (k)^\dagger \hat \Psi(k) \vert \Psi \rangle$ we have:
\begin{equation}
S_k(t) = - \int dk \rho^{(1)}(k,t) \ln [\rho^{(1)}(k,t)].\label{SIE1k}
\end{equation}
The SIEs in Eqs.~\eqref{SIE1x},\eqref{SIE1k} are a measure of the delocalization of the corresponding distributions $\rho^{(1)}(x)$ and $\rho^{(1)}(k)$. Refs.~\cite{Massen1,Massen2,Massen3} establish a universal relation between entropy and the number of interacting particles for diverse systems like atoms, nuclei, and atomic clusters.
Since the distributions in Eqs.~\eqref{SIE1x},\eqref{SIE1k} are relate to the one-body density, they are insensitive to correlations that may be present in the state $\vert \Psi \rangle$~\cite{PNAS,PV,Tsatsos2015,Brezinova2012}.
We can, however, formulate an SIE using the two-body density distributions $\rho^{(2)}(x_1,x_2)=\langle \Psi \vert \hat\Psi(x_1)^\dagger \hat\Psi(x_2)^\dagger \hat \Psi(x_1) \hat{\Psi(x_2)} \vert \Psi \rangle$ and $\rho^{(2)}(k_1,k_2)=\langle \Psi \vert \hat\Psi(k_1)^\dagger \hat\Psi(k_2)^\dagger \hat\Psi(k_1)\hat\Psi(k_2) \vert \Psi \rangle$:
\begin{equation}
S_{\rho-x}(t) = - \int dx_1 dx_2 \rho^{(2)}(x_1,x_2;t) \ln [\rho^{(2)}(x_1,x_2;t)],
\label{SIE2x}
\end{equation}
and
\begin{equation}
S_{\rho-k}(t) = - \int dk_1 dk_2 \rho^{(2)}(k_1,k_2;t) \ln [\rho^{(2)}(k_1,k_2;t)]
\label{SIE2k}
\end{equation}
Here $\rho^{(2)}(x_1,x_2)$ [$\rho^{(2)}(k_1,k_2;t)$] is the diagonal part of the two-body reduced density matrix in position (momentum) space. Since the SIEs in Eqs.~\eqref{SIE2x},\eqref{SIE2k} are computed from two-body quantities, they are measures sensitive to correlations in the many-body state $\vert \Psi \rangle$. By comparing the SIE based on the one-body density [Eqs.~\eqref{SIE1x} and \eqref{SIE1k}] with the SIE based on the two-body density [Eqs.~\eqref{SIE2x},\eqref{SIE2k}] the presence of correlations in the state $\vert \Psi \rangle$ can be inferred.

Since the state we consider is expanded in the MCTDHB theory as $\vert \Psi \rangle = \sum_{\vec{n}} C_{\vec{n}} (t) \vert \vec{n};t \rangle$, we can define an alternative SIE using the coefficients $C_{\vec{n}}$ that characterize the distribution of the state $\vert \Psi \rangle$ in the underlying Fock space [cf. Eq.~\eqref{ansatz}]: 
\begin{equation}
S_c(t) = -\sum_{\vec{n}} |C_{\vec{n}}(t)|^2  \ln \left[|C_{\vec{n}}(t)|^2\right]. \label{SIEC}
\end{equation}
We term this entropy \emph{coefficient Shannon information entropy} (C-SIE), or simply coefficient entropy.
A mean-field state is a single-configuration state, i.e. only a single coefficient contributes in Eq.~\eqref{ansatz}. For such a state $S_c(t)=0$ holds at all $t$. Coefficient entropy $S_c$ thus cannot be produced in a mean-field theory. 
When the state $\vert \Psi \rangle$ spreads across several configurations $\vert \vec{n} ; t \rangle$, several expansion coefficients contribute [cf. Eq.~\eqref{ansatz}] and the coefficient entropy $S_{c}$ gradually increases. 
In the limiting case, when the complete $N$-body Fock space is populated by the state $\vert \Psi \rangle$, all coefficients are equally large and $S_c$ saturates to its maximal value. 

Last, we define an SIE measure related to the emergence of fragmentation \cite{Brezinova2012,Tsatsos2015,SIE_1}, i.e., the emergence of multiple significant eigenvalues $n_i;\;i=1,\dots,M$ of the reduced one-body density matrix of the state $\vert \Psi \rangle$~\cite{Oliver}. These eigenvalues are also referred to as natural occupations and thus we term the following measure of entropy the occupation Shannon information entropy (O-SIE):
\begin{equation}
S_n = - \sum_i^M n_i \ln n_i .
\label{SIEO}
\end{equation}
For a state described by a single-orbital mean-field theory, the reduced density matrix is characterized by only a single eigenvalue and hence $S_n=0$ holds. For single-configuration states with multiple contributing orbitals as well as multiconfigurational states, there maybe several occupation numbers and hence $S_n\neq 0$.
For the condensed gas only a single occupation dominates and the occupation entropy is zero. For an increase in interaction strength, the O-SIE gradually increases and saturates only for a maximally fragmented state.  

\paragraph{\bf Correlation functions:}
The normalized $p$-th order correlation function is defined by 
\begin{equation}
 g^{(p)}(x_1^{\prime},...,x_p^{\prime},x_1, ...,x_p)= 
\frac{\rho^{(p)}(x_1,...,x_p \vert x_1^{\prime},...,x_p^{\prime})}
{\sqrt{\prod_{i=1}^p\rho^{(1)}(x_i \vert x_i )\rho^{(1)}(x_i^{\prime} \vert x_i^{\prime} )}}  
\end{equation}
and is the key quantity to define spatial $p$-th order coherence. Here, $\rho^{(p)}(x_1,...,x_p\vert x_1^{\prime},...,x_p^{\prime}; t)$ is the $p$-th order reduced density matrix of the state $\vert \Psi \rangle$~\cite{RJG}.
In the case of $\vert g^{(p)}(x_1...,x_p,x_1 ...,x_p;t) \vert > 1$ ($<1$), the detection probabilities of $p$ particles at positions $x_1,...,x_p$ are referred to as (anti-)correlated.
Recent progress in experiments in quantum gases has been remarkable and the measurement of higher-order correlation functions is now possible \cite{Hodgman2011,Dall2013,Langen2015,Schweigler2017,Tsatsos2017}. In particular, in the work of Ref.~\cite{Schweigler2017} it has been explicitly shown how a many-body system is characterized via its higher-order correlations.

\section{Results}\label{results}

In this section, we display our numerical results for the SIE measures and spatial correlations in the eigenstates of the Hamiltonian of Eq.~\eqref{eq:HAM} as a function of the strength of repulsive interactions, $\lambda$ . We choose three different sets of system parameters representative for the three distinct phases that the system may be in. Results for the superfluid phase, the Mott insulating phase, and the fermionized phase are discussed in the following Sections~\ref{SF}, \ref{MI}, and \ref{FERM}, respectively.

We define the filling factor as the ratio of the number of atoms $N$ and the number of lattice sites $W$: $\nu= \frac{N}{W}$. For the entire manuscript we focus on commensurate filling factors $\nu=1, 2$ and $7$. We keep the lattice depth $V_0 =12.0$ and lattice site number $d=3$ fixed. We thus change the number of particles to change the filling factor and the interaction strength is gradually tuned to cover all the emergent phases for each filling factor. The transitions between the emergent phases are discriminated using information entropies and correlation functions. 

We choose to present the results in the order of increasing complexity of the found many-body state: $\nu=7,1,2$. Indeed, we find that the $\nu=7$ case is captured within a mean-field theory for the range of the interparticle interaction strengths which we investigate. The $\nu=7$ and $\nu=1$ cases can qualitatively be described within the Bose-Hubbard model as there is no structure that forms within sites. Notably, the fermionization that emerges for the $\nu=2$ case at stronger interparticle interactions is a many-body phase that neither of the aforementioned approaches can describe.

\subsection{Commensurate filling factor $\nu=7$: Superfluid state}\label{SF}
\begin{figure}
\includegraphics[angle=270,width=\linewidth]{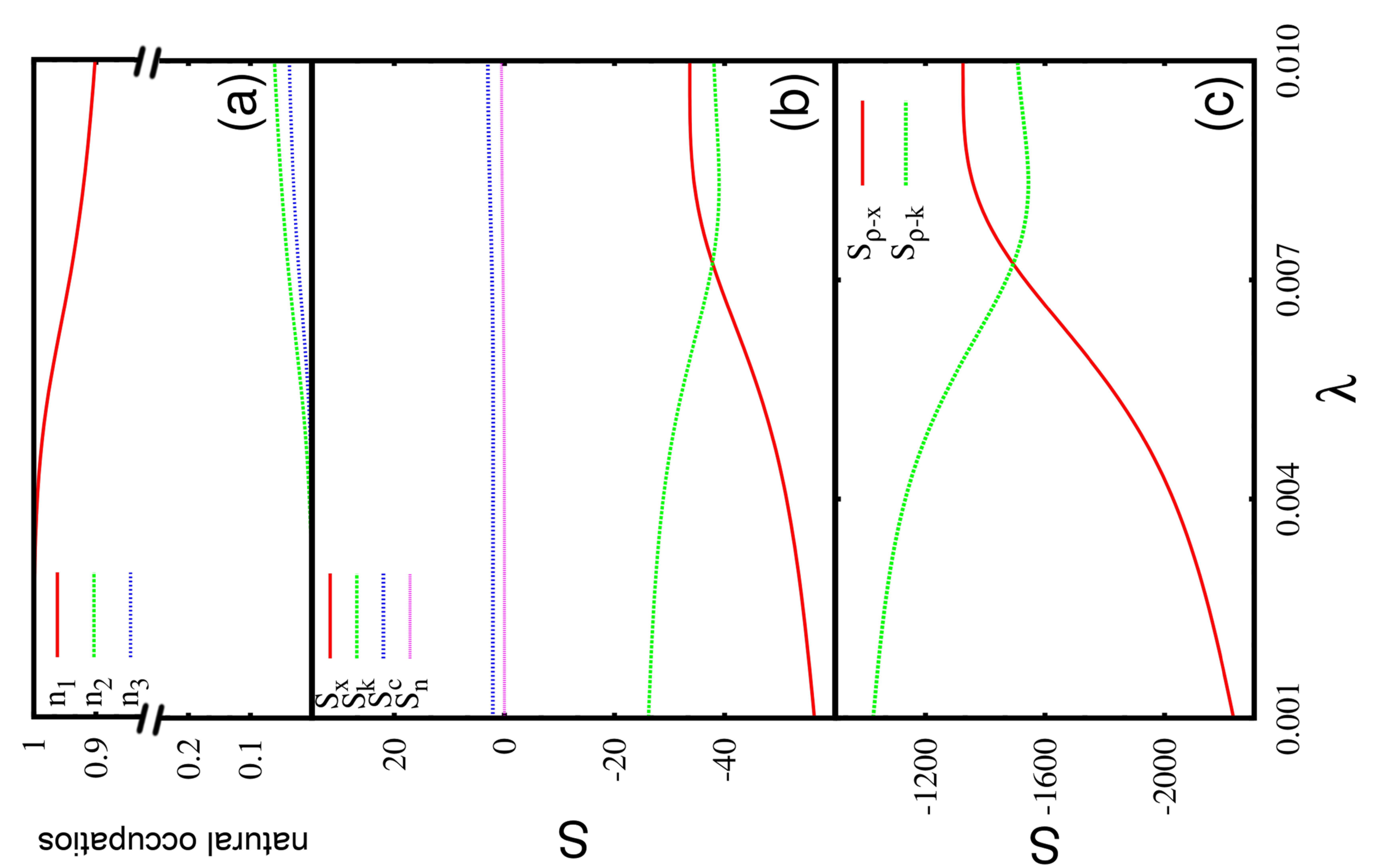}
\caption{{\bf Entropies and occupations in the superfluid phase.} All quantities are shown for $\nu=7$ and $N=21$ bosons as a function of the interaction strength $\lambda$. (a) Eigenvalues of the reduced density matrix -- the ground state is condensed and only a single eigenvalue is macroscopic for the superfluid phase. (b) One-body SIE $S_x$ and $S_k$, C-SIE $S_c$ and O-SIE $S_n$. (c) Two-body SIE $S_{\rho -x}$ and $S_{\rho -k}$. All quantities are dimensionless. See text for discussion.}
\label{fig.fig1}
\end{figure}

Here, we characterize the superfluid phase using natural occupations, normalized correlation functions, and the Shannon information entropy measures that we have introduced. In Fig.~\ref{fig.fig1}(a), we plot the occupations as a function of the interaction strength $\lambda$. For relatively weak interactions, only the first natural orbital is occupied and the population of the second and third orbital remain below $10\%$ for all values of $\lambda<0.01$. In this parameter regime the system is a condensate, according to Penrose and Onsager~\cite{Oliver}, and one can well approximate the many-body wave function with a single-orbital mean-field state $\vert N,0,... \rangle$. To achieve convergence in the occupations, i.e. the eigenvalues of the reduced one-body density matrix the interaction strengths that we consider, three orbitals [$M=3$, cf. Eq.~\eqref{ansatz}] are enough; adding more orbitals does not quantitatively change the many-body state.

We plot the one-body SIE for the density and momentum distributions as a function of the interaction strength $\lambda$ in Fig.~\ref{fig.fig1}(b). As $\lambda$ gradually increases the (momentum) density within each well is broadened (narrowed) and consequently the SIE computed from the density (momentum) distribution, $S_{x}$ ($S_{k}$) [see Eqs.~\eqref{SIE1x} and \eqref{SIE1k}] increases (decreases). Note that the curves for $S_x$ and $S_k$ cross each other for $\lambda =0.007$. In Fig.~\ref{fig.fig1}(b), we also present the C-SIE $S_{c}$ and O-SIE $S_{n}$. As anticipated, the values of $S_{c}$ and $S_{n}$ are very close to zero, because the condensate is phase coherent and can be well described by a macroscopically occupied single-particle state and the Gross-Pitaevskii mean-field equation yields a good description. 

In Fig.\ref{fig.fig1}(c), we plot the two-body SIE [Eqs.~\eqref{SIE2x},\eqref{SIE2k}]. The two-body SIEs behave similarly to the one-body entropy measures; the crossing of the momentum-based and the real-space-based SIE happens at the same point $\lambda =0.007$. However the rate of increase in $S_{\rho -x}$ is greater than that in $S_x$ and, likewise, the rate of decrease of $S_{\rho -k}$ is greater than that in $S_k$. A similar trend was observed in the calculation of one-electron and two-electron entropies in Ref.~\cite{Guevara1}. The fact that the two-body density matrix can -- to a good approximation -- be written as a product of one-body densities for the coherent superfluid gas is a possible explanation for this behavior.

\begin{figure}
\includegraphics[angle=270,width=\linewidth]{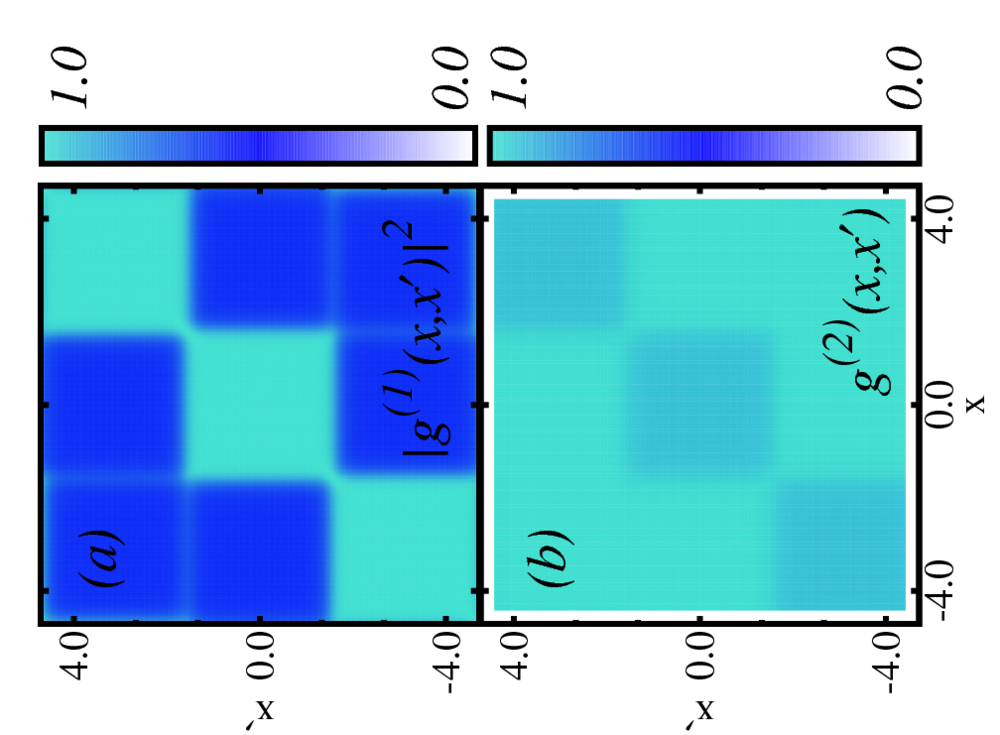}
\caption{{\bf (a) first- and (b) second-order normalized correlation function for the superfluid phase, i.e. $\nu=7, N=21$.} The interaction strength is $\lambda=0.01$ while the rest of the parameters are as in Fig.~\ref{fig.fig1}. (a) Complete first-order coherence is observed within each well;  $\vert g^{(1)} \vert \approx 1$ for all $x\approx x'$. Significant inter-well coherence can be inferred from $\vert g^{(1)} \vert \in [0.5,1]$ for all $x\neq x'$ on the off-diagonal. (b) Second-order coherence is retained among distinct wells $g^{(2)}(x,x')\approx 1$ for all $x\approx x'$. The fact that $g^{(2)}$ is maximal among distinct wells but drops within each well indicates that the probability for detecting two particles in different wells is higher than that of finding them in the same well. This behavior is a consequence of the interparticle interactions and it is termed antibunching. All quantities are dimensionless. See text for discussion.}
\label{fig.fig2}
\end{figure}

In Fig.~\ref{fig.fig2}(a), we plot the absolute value of the normalized first-order correlation function, $\vert g^{(1)}(x^{\prime},x)\vert^2$ for $\lambda= 0.01$. We see that in the $(x,x')$ region where the density is localized $\vert g^{(1)}\vert^2 \approx 1$ holds. We infer that coherence within and between sites is maintained. Fig.~\ref{fig.fig2}(b) shows the two-body correlation function $g^{(2)}(x^{\prime},x,x^{\prime},x)\equiv g^{(2)}(x,x^{\prime})$ for the same parameters as in Fig.~\ref{fig.fig2}(a). We find $g^{(2)}(x,x')$ to be close to unity for all $x',x$. Remarkably, second-order coherence between different wells is perfectly maintained: $g^{(2)}(x,x')\approx 1$ for the off-diagonal $x'\neq{}x$ with $\vert x'-x \vert$ larger than the size of a single well. 
The diagonal part of the two-body correlation function is slightly depleted, i.e. $g^{(2)}(x,x')\lesssim 1$ for the diagonal $x' \sim x$ if $\vert x'-x \vert$ is smaller than the size of a single well. This signifies that anti-bunching starts to develop for particles within the same site due to the repulsive interactions: second order coherence is locally decreased, even though the interaction strength $\lambda$ is relatively small.

\subsection{Commensurate filling factor $\nu=1$: transition from condensation to fragmentation}\label{MI}
\begin{figure}
\includegraphics[angle=270,width=\linewidth]{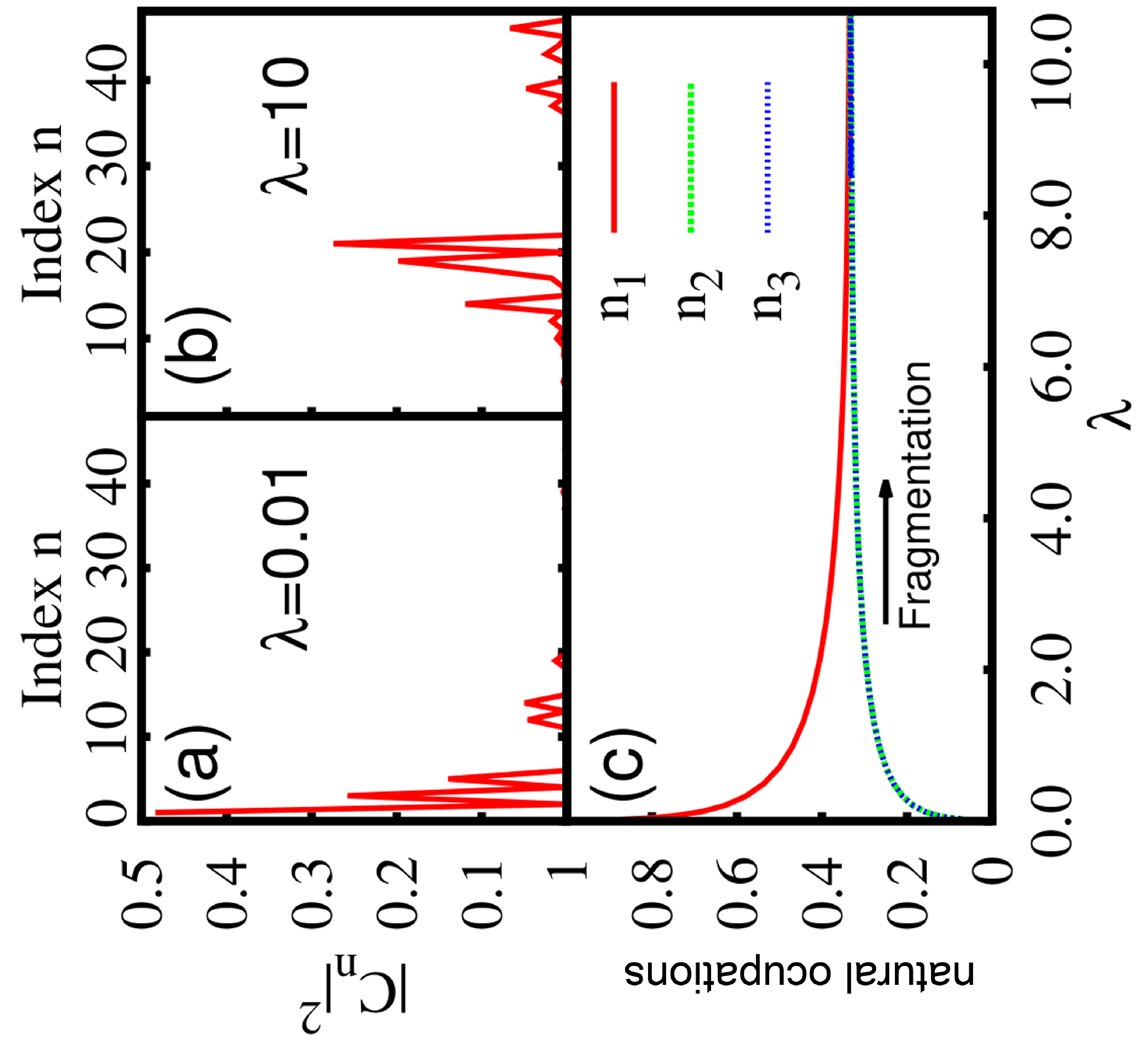}\\  
\caption{{\bf Localization and delocalization in Fock space in the superfluid to Mott-insulator transition at $\nu=1,N=3$.} Panels (a) and (b) show the distribution of the magnitude of the coefficients $\vert C_{n}\vert^{2}$ as a function of the index $n$ for the superfluid at interaction strength $\lambda=0.01$ in (a) and the Mott insulator at interaction strength $\lambda=10.0$ in (b). The index $n$ is computed from the vector $\vec{n}$ using the mapping described in Ref.~\cite{MAP}. In (a) the $\vec N,0,...\rangle$ coefficient and its neighbors dominate while all others are small; the state is \emph{localized} in the many-body Fock space. At (b) the larger interactions force the many-body state to spread at a larger part of the available space and many coefficients are significant; the state is \emph{delocalized} in the many-body Fock space. Panel (c): Populations of the first three natural orbitals. As $\lambda$ increases, the occupation of the first orbital gradually decreases while another two orbitals begin to contribute (green and blue curves). Eventually the state becomes three-fold fragmented ($n_1=n_2=n_3=1/3$) at $\lambda\gtrsim6.0$. All quantities are dimensionless.}
\label{fig.fig3}
\end{figure}

We now turn our attention to the Mott insulating phase and investigate its many-body physics from the viewpoint of correlation functions, coherence and Shannon information entropies. To obtain an archetypical Mott-insulator for strong interactions, we consider $N=3$ bosons in three wells, i.e. a single atom per site. To quantify the departure of the many-body state $\vert \Psi \rangle$ from a mean-field state in the superfluid-to-Mott-insulator transition, we plot the expansion coefficients $\vert C_{n} \vert^{2}$ [Eq.~\eqref{ansatz}] as a function of the index $n$ of the basis states for two choices for the interaction strength: $\lambda=0.01$ which puts the system onto a superfluid phase and $\lambda=10.0$ that renders it a Mott-insulator [see Fig.~\ref{fig.fig3}(a) and (b)].
In the superfluid phase ($\lambda=0.01$) the number of significant coefficients is only a small portion of the total number of configurations $N_{conf}=\binom{N+M-1}{N}$. We refer to such a state as localized. We emphasize that our use of the term ``localized'' refers to many-boson Fock space and not real-space; localized states are thus close to a mean-field description where only a single coefficient would contribute. However, as $\lambda$ increases a larger amount of coefficients become significant; we refer to such a state as delocalized [cf. $\lambda=10.0$ in Fig.~\ref{fig.fig3}(b)]. 

In Fig. \ref{fig.fig3}(c), we plot the corresponding occupation of the first, second, and third natural orbital as a function of the interaction strength. With an increase of $\lambda$, the occupation of the first natural orbital gradually decreases while the occupations of the second and third natural orbitals gradually increases. For large interactions, the first three natural occupations saturate at $33.33\%$. In our computations with $M=6$ orbitals, all but the first three orbitals had occupations smaller than $10^{-7}$. 
Three-fold fragmentation in a triple well indicates loss of partial coherence of distinct sites and is a signature for the transition to the insulating phase \cite{Streltsov2011}. Thus, across the superfluid-to-Mott-insulator transition the increase of the interaction strength drives the state from a condensed to a fragmented one. Generally, in the Mott-insulating phase there are as many significant eigenvalues of the reduced density matrix as there are lattice sites.

\begin{figure}
\includegraphics[angle=270,width=\linewidth]{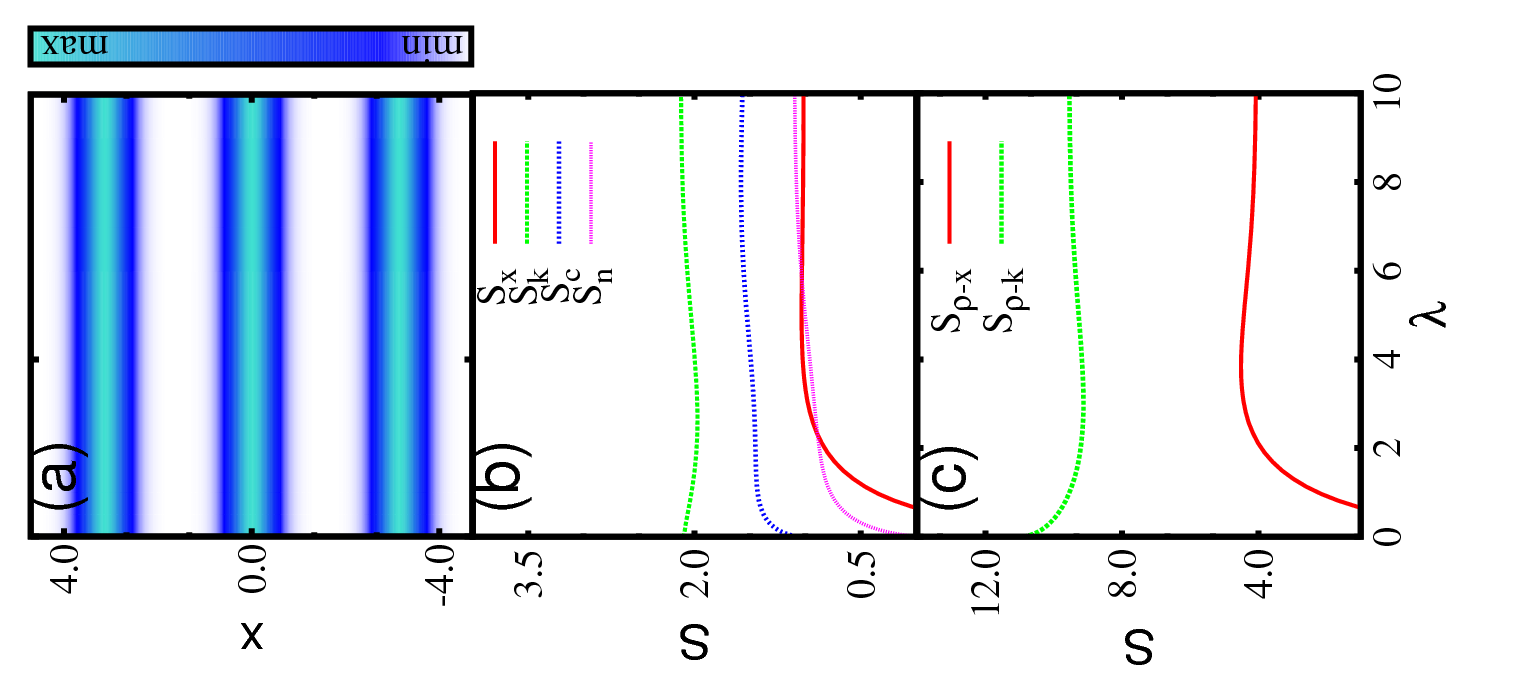}\\  
\caption{{\bf One-body density and entropies in the superfluid-to-Mott-insulator transition at unit filling $\nu=1$ and $N=3$ as a function of the interaction strength.} (a) The one-body density as a function of interaction strength $\lambda$ shows almost no variation in the transition from a condensed superfluid to a fragmented Mott insulator. $(b)$: Shannon information entropy of the one-body (momentum) density $S_x$ ($S_k$), the occupation entropy $S_n$ and the coefficient entropy $S_c$ all as a function of $\lambda$. 
(c) SIE of the reduced two-body (momentum) density matrix $S_{\rho-x}$ ($S_{\rho -k}$). For the SIE measures in (b) and (c), the curved part for weak interactions signifies the transition from a condensed superfluid to a fragmented Mott insulator. Within the Mott-insulating phase, for larger interaction strengths $\lambda$, all SIE measures saturate. All quantities are dimensionless, see text for further discussion.}
\label{fig.fig4}
\end{figure}

From Fig.~\ref{fig.fig4}(a) we see that the above transition cannot be inferred from the density alone. Indeed, the density before and after the transition (in the $\lambda$ space) are identical and many-body measures, such as the entropies, are required to mark the transition. Examining the C-SIE $S_{c}$ and O-SIE $S_{n}$ we observe that for small  $\lambda$ both start at zero, as only a single coefficient contributes to the the state, which is a condensed superfluid. The behavior of the SIE measures computed from the one- and two-body density are in mutual agreement with the C-SIE and O-SIE measures, compare Figs.~\ref{fig.fig3} and Fig.~\ref{fig.fig4}(b),(c). As $\lambda$ rises, the C-SIE and O-SIE measures increase and the many-body state converges to a pure Mott-insulator while fragmentation, as well as all SIE measures saturate.

We shall now explore this saturation of entropies using an analogous Gaussian Orthogonal Ensemble (GOE). The GOE of random matrices can be formulated for isolated quantum systems of interacting particles which exhibit time reversal symmetry and rotational invariance~\cite{rigo,sant3,rigol}. The entropies evaluated for the GOE provide estimates for the entropies of many-body systems for the limiting case of maximal disorder. The GOE of random matrices with infinite interaction has entropy $S_{c}^{GOE}=\ln (0.48 D_c)$, where $D_c \times D_c$ is the dimension of the considered random matrices.

In the present $\nu=1$ case, three bosons are distributed in (essentially) three orbitals, $N_{orb}=3$ [Fig.~\ref{fig.fig3}(c)]. Hence, the size of the GOE is $D_c= \binom{N + N_{orb} -1}{N} = 10$ thus obtaining $S_{c}^{GOE} = 1.568$. This is in excellent agreement with the value at which our numerical result for the C-SIE $S_c$ saturates: $1.552$. The small discrepancy is because the interaction strength is still finite. The coefficient entropy thus saturates at the value predicted from the GOE for sufficiently large $\lambda$. 
For the O-SIE, $S_n$, the dimensionality of the GOE $D_n=3$ is equal to the number of occupied orbitals in our MCTDHB treatment and $S_{n}^{GOE} = -\sum_{i=1}^{D_n}\frac{1}{D_n} \ln\left(\frac{1}{D_n}\right) = \ln (D_n) = 1.098$. Our numerical result for the saturated value of the O-SIE is identical, $S_{n} = 1.098$ \cite{M9comparison}.

\begin{figure}
\includegraphics[angle=270,width=\linewidth]{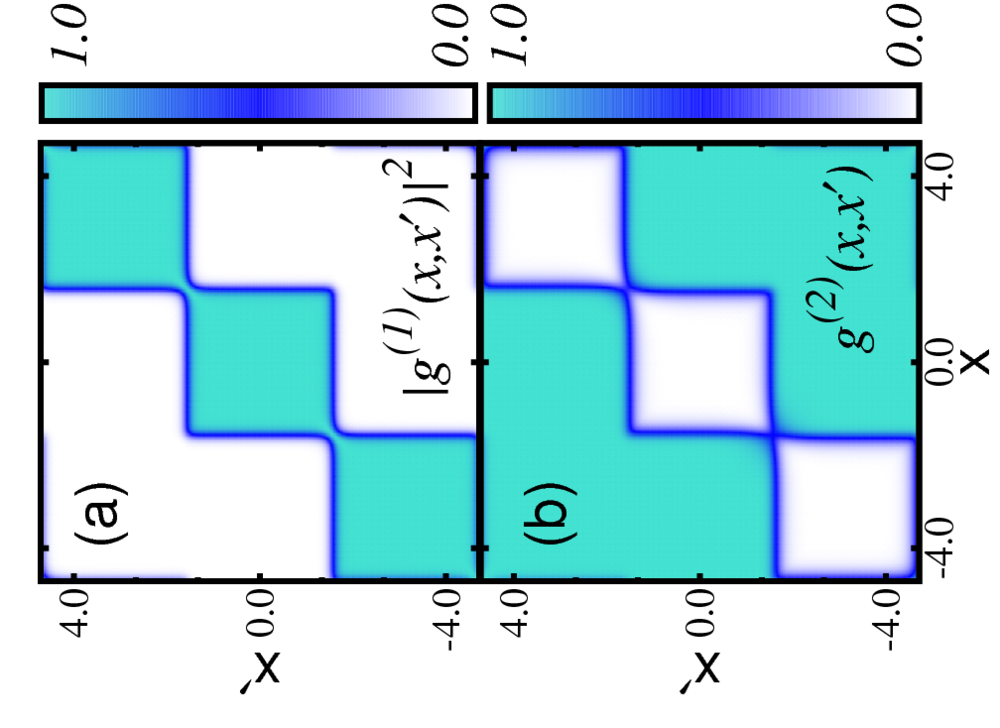}\\  
\caption{{\bf First-order and second-order coherence of a fragmented Mott-insulator at $\nu=1, N=3$.} (a) The normalized first-order correlation function $\vert g^{(1)}(x,x')\vert^{2}$ for $\lambda=10$ exhibits three separated regions with $\vert g^{(1)} \vert^2\approx1$ along the diagonal and $\vert g^{(1)} \vert^2\approx0$ on the off-diagonal. Thus, the first-order coherence is maintained within wells and lost between them. (b) The normalized second-order correlation function $g^{(2)}(x,x')$ also for $\lambda=10$. The ``correlation hole'' is clearly seen and $g^{(2)}\approx 0$ holds for the diagonal $x\approx x'$; two particles are thus almost never simultaneously found in the same well. Second-order coherence $g^{(2)} \approx 1$ on the off-diagonal signifies that, in a measurement, two uncorrelated particles will almost surely be detected in distinct wells. The Mott-insulating phase is characterized by independent lattice sites with strongly localized particles. All quantities are dimensionless.}
\label{fig.fig5}
\end{figure}

We analyze the first-order and second-order coherence in the fragmented Mott insulator for $\lambda = 10.0$ in Fig.~\ref{fig.fig5}. The diagonal of the first-order correlation function shows three completely separated coherent regions where $\vert g^{(1)} \vert^2\approx1$; the coherence of bosons within the same well is maintained while it is lost between distinct wells, since $ \vert g^{(1)} \vert^2\approx 0$ at the off-diagonal, as expected for the fully localized particles in a Mott insulator. 
Looking at $g^{(2)}$ we see that coherence is maintained ($g^{(2)}\approx 1$) at the off-diagonal but not at the diagonal. 
The vanishing diagonal part of the normalized two-body correlation function -- referred to as the correlation hole~\cite{Brouzos} -- translates to the fact that the probability of finding a double occupation of a single well is practically zero. Second order coherence between wells is maintained, because the outcome of a two-particle detection is almost always a certain and uncorrelated detection of two particles in distinct wells.

\begin{figure}
\includegraphics[angle=270,width=\linewidth]{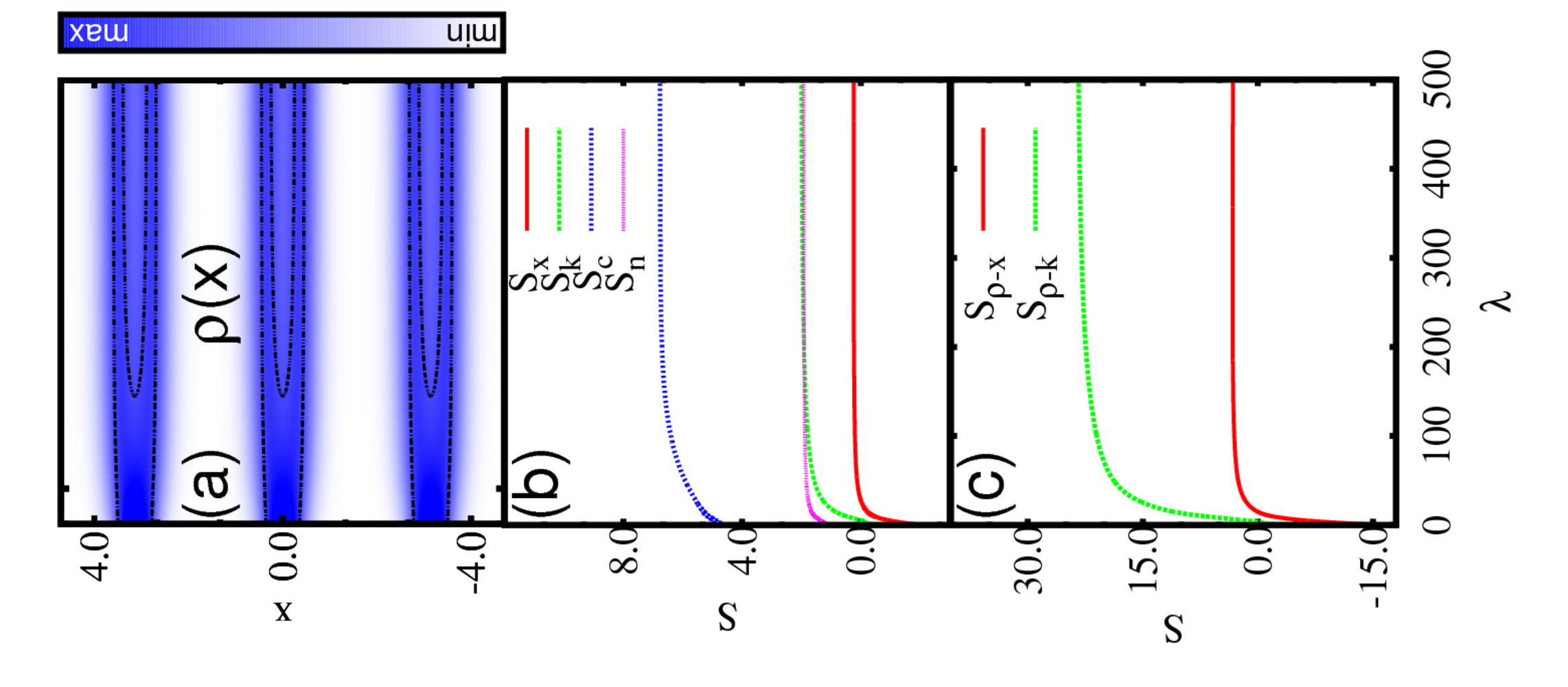}
\vspace{-9mm}
\caption{{\bf One-body density and entropy measures for the transition from the Mott-insulator to fermionization.} All panels are for $\nu=2,N=6$ and are plotted against the interaction strength $\lambda$. (a) Plot of the one-body density; the transition to fermionization is visible in the formation of a two-hump intrawell structure -- for large $\lambda$, the number of peaks in the density is equal to the number of particles $N$. The dashed contour line $\rho(x)=0.22$ is drawn to visualize the two maxima of each well. (b) SIE computed from the one-body (momentum) density $S_x$ ($S_k$), the coefficients entropy $S_c$, and the occupation entropy $S_n$. (c) SIE computed from the two-body (momentum) density, $S_{\rho-x}$ ($S_{\rho-k}$).
Compared to the system with a single boson per well, the SIE measures saturate only for comparatively large interaction strengths [compare panels (b) and (c) to Fig.~\ref{fig.fig4} panels (b) and (c), respectively]. All quantities are dimensionless, see text for further discussion.}
\label{fig.fig6}
\end{figure}

\subsection{Commensurate filling factor $\nu=2$: from a condensed superfluid via a fragmented Mott insulator to fermionization}\label{FERM}
We now consider a system with $\nu=2$ bosons per site and characterize its phases using the density, normalized correlation functions and Shannon information entropy measures. In lattices with commensurate filling factor larger than one, i.e. with more than one boson per site, two phase transitions can be observed as a function of the strength of the interparticle interactions $\lambda$: The superfluid-to-Mott-insulator transition, that is also present if the filling factor is equal to one, emerges for intermediate interaction strengths. For large interaction strengths the Mott-insulator-to-fermionization transition -- absent for $\nu=1$ -- now occurs. Since the physics of the superfluid-to-Mott-insulator transition is unchanged, we focus here on the Mott-insulator-to-fermionization transition. The pathway from the condensed superfluid via the fragmented Mott insulator to the fermionization regime depends on the shape of the trapping potential~\cite{Ofir2}. The optical lattice with filling factor two is the generic system to for this transition to be studied. 

Similarly to Sec.~\ref{MI}, we see that the system transitions to MI at $\lambda\approx10$. The transition cannot be seen in the one-body density, since the latter remains unchanged across the SF-to-MI transition as $\lambda$ increases [see Figs.~\ref{fig.fig6}(a) and \ref{fig.fig4}(a)]. 
When the interaction strength $\lambda$ is sufficiently large ($\lambda\gtrsim200$), the Mott-insulator-to-fermionization transition takes place and the number of maxima in the one-body density becomes equal to the number of bosons of the system, see Fig.~\ref{fig.fig6}(a). In this case the bosonic distribution resembles a fermionic one where Pauli exclusion forbids the spatial overlap of the particles. 
The transition to fermionization is thus manifest in the formation of intrawell structure in the density. Such intrawell structures cannot be appropriately described by the single-band Hubbard model. When a complete set of Wannier states is considered in each site, i.e. a multi-band Hubbard model, an intractably large number of bands quickly becomes necessary~\cite{optimal_lattices}. Thus, the transition to fermionization is a hallmark for the inapplicability of Hubbard models.

\begin{figure}
\includegraphics[angle=270,width=\linewidth]{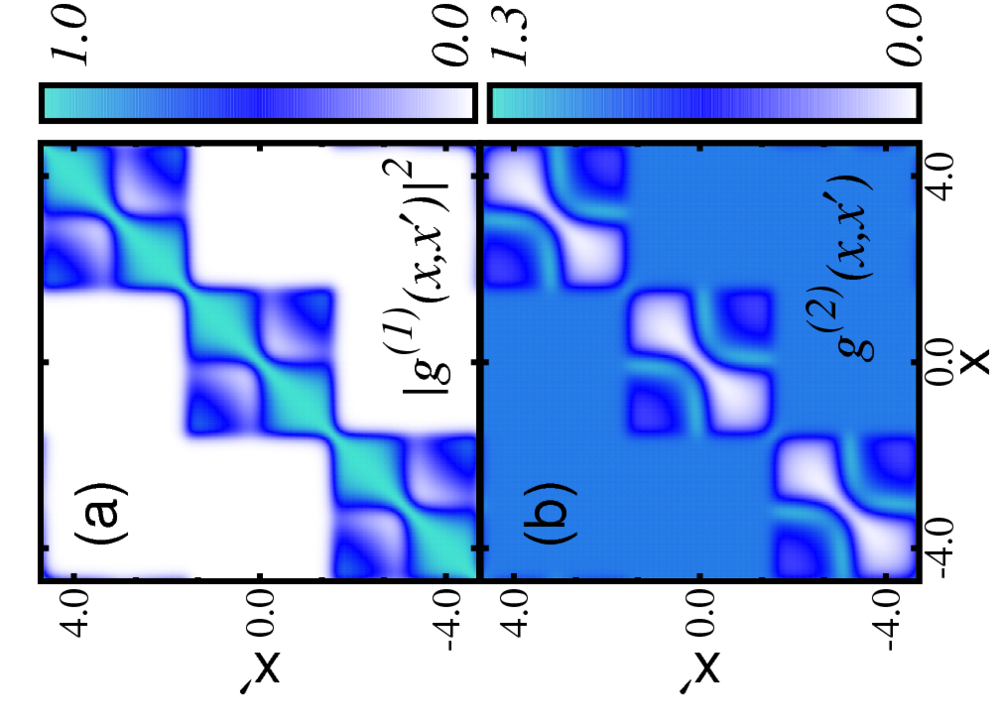}\\  
\caption{{\bf First- and second-order coherence for $N=6$ fermionized bosons and $\nu=2$.} Both panels show normalized correlation functions for an interaction strength $\lambda=200$. (a) The normalized first-order correlation function $\vert g^{(1)}(x,x')\vert^{2}$. The coherence between wells is completely lost, i.e. $\vert g^{(1)}\vert^{2}\approx 0$ for the off-diagonal. The fermionization of the system is manifest in the loss of first-order coherence also within each of the wells [cf. Fig.~\ref{fig.fig5}(a)]. (b) The normalized second-order correlation function $g^{(2)}(x,x')$ has a fully developed correlation hole, $\vert g^{(2)}\approx0$ on the diagonal. No bunching or anti-bunching is seen between particles in distinct wells, i.e.$g^{(2)}\approx1$ on the off-diagonal. Fermionization is manifested in the fact that $g^{(2)}<1$ as long as $x,x'$ are within each of the wells [cf. Fig.~\ref{fig.fig5} (b)]. All quantities shown are dimensionless, see text for further discussion.}
\label{fig.fig7}
\end{figure}

The SIE measures plotted in Fig.~\ref{fig.fig6}(b) and \ref{fig.fig6}(c) as a function of $\lambda$ also saturate in the transition to fermionization. The saturation of SIEs in the Mott-insulator-to-fermionization transition with two bosons in each well, however, takes place at a much larger interaction strength ($\lambda\sim200$) as compared to the saturation of SIEs for the superfluid-to-Mott-insulator transition ($\lambda\sim10$) in the case of $\nu=1$. Interestingly, no GOE analogue for the fermionized state has been established so far and, consequently, we cannot provide a comparison the values that the entropies shall saturate to.

We plot the normalized one-body and two-body correlation functions $g^{(1)}$ and $g^{(2)}$ to determine how the Mott-insulator-to-fermionization transition reflects in the coherence of the many-body state, see Fig.~\ref{fig.fig7}.
The first and second order coherence between different wells is similar for the fragmented and the fermionized Mott insulator, i.e. the off-diagonal parts of the normalized correlation functions $g^{(1)}$ and $g^{(2)}$ are practically identical (cf. Figs.~\ref{fig.fig5} and \ref{fig.fig7}).
However, in contrast to the correlation functions of the fragmented Mott insulator with a single boson per site, the correlation functions for the fermionized Mott insulator feature a distinct intrawell structure (cf. Figs.~\ref{fig.fig5} and ~\ref{fig.fig7}).

\section{Conclusions}\label{Conclusion}
We have shown that various information entropies can be defined to reflect and quantify many-body properties of the ground state of interacting bosons in one-dimensional lattices. These entropies serve the dual purpose to $i)$ identify phase transitions not seen in the single-particle properties of the system and $ii)$ allow a connection to known results from random matrix theory. 

We demonstrated that our SIE measures are in agreement with predictions for the Gaussian orthogonal ensemble (GOE) of random matrix theory in the case of strong interactions. For sufficiently strong interactions and a single boson per site the system is fully in the Mott insulating regime and all our SIE measures saturate to their respective GOE values. For sufficiently strong interactions and two bosons per site the system is fermionized and the SIE measures are saturated. For this case, however no GOE estimation is available.

All the emergent phases -- the superfluid, the Mott-insulator, and the fermionized state -- are identified with a distinct value of their SIE measures. Our analysis of the SIE measures is complemented by an analysis of the density and the spatial first- and second-order coherence. We showed that all emergent phases have characteristic density, correlation, and coherence patterns. 
We hence demonstrated that the superfluid, the Mott-insulating and the fermionized phases of bosons in one-dimensional lattices are not only characterized uniquely by their densities and Glauber normalized correlation functions~\cite{RJG,Ofir1}, but also by their SIE values. This corroborates the fundamental connection between correlation functions and entropy measures conjectured already in Refs.~\cite{SIE_1,Brezinova2012}.

In the transition to fermionization, intrawell structure in the densities and correlations of the many-body state is formed; this marks the breakdown of the single-band Hubbard model. Since the SIE measures and Glauber normalized correlations allow to identify the fermionized state, we also demonstrated two independent ways to assess the applicability of the Hubbard description.

Studying higher-order correlations and their `holes' in the $(x,x')$ would consist a natural extension of our works. Moreover, phases beyond fermionization and the relation of our findings for the stationary states to quench dynamics are open questions.

\begin{acknowledgments}
RR acknowledges UFC fellowship and B. Chakrabarti acknowledges FAPESP (grant No. 2016/19622-0). AG and MCT acknowledge FAPESP and AG thanks CNPq for financial support. B. Chatterjee acknowledges financial support from the Department of Science and Technology, Government of India under the DST Inspire Faculty fellowship. AUJL acknowledges financial support by the Austrian Science Foundation (FWF) under grant No. F65 (SFB ``Complexity in PDEs''), and the Wiener Wissenschafts- und TechnologieFonds (WWTF) project No MA16-066 (``SEQUEX'').
\end{acknowledgments}


\begin{thebibliography}{References}
\bibitem{M9comparison} We repeated the calculations for $N=6, M=9$ and obtained $D_c=3003, D_n=9$ and $S_c=6.764$ and $S_n=1.925$ for the saturated entropy values. For comparison, we found $S_c^{GOE}=7.273$ and $S_n^{GOE}=2.197$.
\end{thebibliography}
\end{document}